\documentclass{PoS}

\title{Can the James Webb Space Telescope detect isolated population III stars?}

\ShortTitle{Can the James Webb Space Telescope detect isolated population III stars?}

\author{\speaker{Claes-Erik Rydberg}\\
Department of Astronomy, Stockholms University\\
E-mail: \email{Claes-Erik.Rydberg@astro.su.se}}

\author{Erik Zackrisson\\
Department of Astronomy, Stockholms University\\
E-mail: \email{ez@astro.su.se}}

\author{Pat Scott\\
Department of Physics, McGill University\\
E-mail: \email{patscott@physics.mcgill.ca}}

\abstract{Isolated population III stars are postulated to exist at approximately z=10-30 and may attain masses up to a few hundred solar masses. The James Webb Space telescope (JWST) is the next large space based infrared telescope and is scheduled for launch in 2014. Using a 6.5 meter primary mirror, it will probably be able to detect some of the first galaxies forming in the early Universe. A natural question is whether it will also be able to see any isolated population III stars. Here, we calculate the apparent broadband AB-magnitudes for 300 M$_{\odot}$ population III stars in JWST filters at z=10-20. Our calculations are based on realistic stellar atmospheres and take into account the potential flux contribution from the surrounding HII region. The gravitational magnification boost achieved when pointing JWST through a foreground galaxy cluster is also considered. Using this machinery, we derive the conditions required for JWST to be able to detect population III stars in isolation. We find that a detection of individual population III stars with JWST is unlikely at these redshifts. However, the main problem is not necessarily that these stars are too faint, once gravitational lensing is taken into account, but that their surface number densities are too low.}

\FullConference{Cosmic Radiation Fields: Sources in the early Universe\\
November 9-12, 2010\\
Desy, Germany}

\begin{document}

\section{Introduction}

Both theoretical arguments and numerical simulations, reviewed in \cite{2004ARA&A..42...79B}, strongly support the notion that population III stars were very massive, significantly more so than the population I and population II stars that formed later on. When stars form in metal enriched gas, the Jeans mass is lower, which leads to fragmentation and thus lower masses. Lacking metals, a chemically unenriched cloud does not fragment in the same way and higher stellar masses are therefore possible. It has been argued that two different classes may have existed: population III.1 and population III.2. Population III.1 stars formed in dark matter minihalos of mass 10$^5$-10$^6$ M$_{\odot}$ at z=10-30. As described in \cite{2009Natur.459...49B}, probably only one star, with an average mass of $\sim$ 100 M$_{\odot}$ were produced in every subhalo. This follows as the massive star that is formed emits a lot of UV radiation which destroys the molecular hydrogen in the parent cloud, preventing further cooling and star formation. Population III.2 stars then form due to HD cooling with an average mass of $\sim$ 10 M$_{\odot}$. Even though recent simulations indicate a more complex scenario, see \cite{2010arXiv1006.1508C} and \cite{2008ASPC..393..275S}, we will for simplicity assume that only one heavy population III.1 star is formed per halo. As we will argue, a flux boost due to gravitational lensing by a foreground galaxy cluster will be necessary to detect such stars. To assess the required properties of the lens in order to see on average one population III.1 star per survey field, we will in general use the most optimistic parameters for the calculations. Although debatable, it has become standard practice to assume a mass range of 60 M$_{\odot}$ to 300 M$_{\odot}$ for such stars, (e.g. \cite{2009Natur.459...49B} and \cite{2009ApJ...694..879T}). Here, we will explore the most optimistic scenario possible, and therefore adopt 300 M$_{\odot}$ for our population III.1 stars.

The spectral energy density of a primordial star has previously been investigated by Bromm et al \cite{2001ApJ...552..464B}, with the result that isolated population III stars were deemed to be too faint for detection with the James Webb Space telescope (\cite{2006SSRv..123..485G} and \cite{2009MNRAS.399..639G}). Even though the possibility of gravitational lensing has been briefly mentioned in \cite{2006SSRv..123..485G} it has not been investigated in detail. However, gravitational lensing seems to offer the only plausible route to direct detection of isolated population III stars before they explode as supernovae (e.g. \cite{2002luml.conf..369H}).

\section{Modelling the spectra of Population III stars}

The population III properties derived in \cite{2002A&A...382...28S} have been used as the basis for our magnitude calculations. Here, we adopt the main sequence properties of these stars and thereby neglect the star emitting less ionizing flux when aging. We have used realistic stellar atmospheres and have also taken into account the potential flux contribution from the surrounding HII region. To model the atmosphere we have used the publicly available TLUSTY code, see \cite{1995ApJ...439..875H}. This code computes 1D, non-LTE, plane-parallel stellar model atmospheres and spectra, given a certain input stellar composition, surface temperature and gravity.

The observed flux from a population III star probably originates mostly from the HII region that is created around it. To model this, we have used the publicly available photoionization code Cloudy \cite{1998PASP..110..761F}. We use this to calculate the spectrum from spherically symmetric HII regions around these stars. Both the nebular continuum and emission lines are predicted. This procedure results in an optimistically bright emission spectrum, since real nebulae experience feedback effects that could give rise to holes in the HII region. This exposes the star underneath and the spectrum emerging from the stellar atmosphere. The region could also break out of the gravitational well of the host halo if the feedback is strong enough. As described in \cite{2009ASPC..419..335J}, this dillutes the nebular flux and exposes more of the purely stellar spectrum.

Because of likely Ly$\alpha$ absorption in the neutral IGM at z$>$6, the spectra have both the continuum intensity at wavelengths shorter than the Ly$\alpha$ and the Ly$\alpha$ line set to zero. At high redshift this has a huge impact on the magnitudes since this effect will be redshifted into the filter we are using, as described below.

\section{JWST broadband fluxes of population III stars}
\label{sec:JWSTbroadbandfluxes}

\begin{figure}
    \begin{center}
        \includegraphics[width = 11 cm]{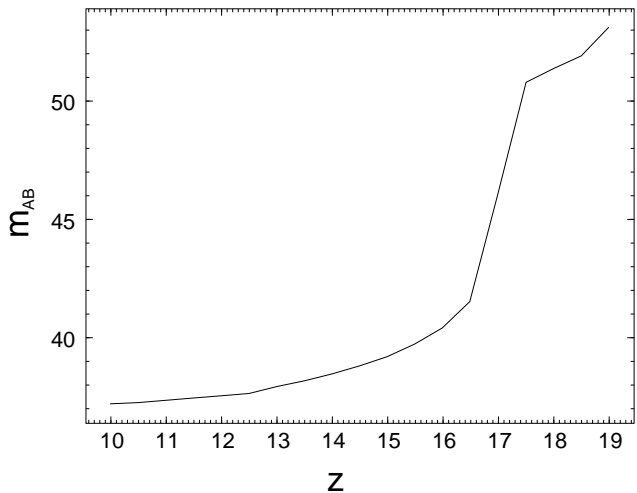}
        \caption{The predicted AB-magnitude in the NIRCam F200W filter of a 300 M$_{\odot}$ population III star as a function of redshift.}
    \label{fig:magnitudes}
    \end{center}
\end{figure}

\begin{figure}
    \begin{center}
        \includegraphics[width = 11 cm]{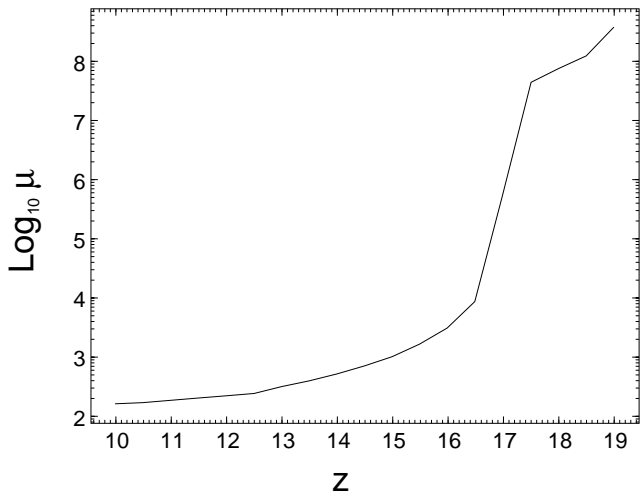}
        \caption{The gravitational magnification required to detect a 300 M$_{\odot}$ population III star through deep imaging with the JWST in the NIRCam F200W filter.}
    \label{fig:necessarymagnification}
    \end{center}
\end{figure}

Figure \ref{fig:magnitudes} shows the predicted AB-magnitude in the JWST NIRCam/F200W filter for our 300 M$_{\odot}$ star. This instrument/filter have been selected to optimize the probability of detecting the star through deep imaging. The NIRCam/F200W filter has its main transmission between 17,500 $\mbox{\AA }$ and 22,500 $\mbox{\AA }$. This means that the redshifted Ly$\alpha$ (1216 $\mbox{\AA}$) absorption starts affecting the magnitudes at approximately z=15. For objects at z=18 it has absorbed most of the spectrum entering the filter. The effect from Ly$\alpha$ absorption can be seen as the sharp increase of magnitudes in this redshift interval. With a very optimistic scenario of a 100 hour exposure, a detection limit of m$_{200}$=31.68 can be reached at 5$\sigma$. Since the flux of a 300 M$_{\odot}$ star is predicted to be m$_{200}$>37 magnitude, a flux boost due to gravitational lensing is required for a detection to be feasible.  The gravitational lens magnification required for this has been plotted as a function of z in figure \ref{fig:necessarymagnification}. At z=10, a magnification of 160 is required to detect this object. At higher redshift, the required magnification becomes even greater.

\begin{figure}
    \begin{center}
        \includegraphics[width = 11 cm]{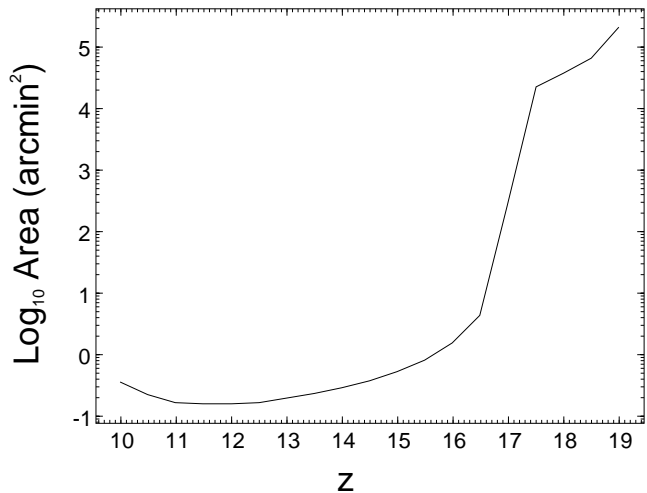}
        \caption{The required lens-plane area over which the magnification would have to be equal to that derived in figure 2, in order to allow on average one 300 M$_{\odot}$ to be detected.}
    \label{fig:neededarea}
    \end{center}
\end{figure}

Using the Trenti and Stiavelli models \cite{2009ApJ...694..879T}, the number of stars per square arcmin and redshift can be calculated. Here, we have adopted the no Lyman-Werner scenario, since this results in the most optimistic prospects for detecting isolated population III stars at z<15. The magnified area required for detecting on average one population III star of 300 M$_{\odot}$ can be seen in figure \ref{fig:neededarea}. In this graph we have also used the very unrealistic assumption that \emph{all} stars attain the mass of 300 M$_{\odot}$. In reality just a fraction will be, and the area should increase accordingly. The area is moreover calculated with the assumption that it has exactly the required magnification. In a real lens, the magnification will generally be higher in certain regions. This means that less area is covered in the source plane, and that a larger area is required in the lens plane. The area actually decreases from z=10 to z=12, where an area of about 0.15 square arcmin is needed, simply because the models predict more stars at this epoch. The area at redshift 12 corresponds to a magnification of 200. For comparison, the gravitational lens with the greatest known Einstein radius (MACS J0717.5+3745 at z=0.546, \cite{2009ApJ...707L.102Z}) has a magnification of 100-300 (average 160) in an area of 0.3 square arcmin at redshift 6-20 (see \cite{2010ApJ...717..257Z}). While this is in the right ballpark to allow 300 M$_{\odot}$ population III stars to be detected, this requires that a number of desperately unlikely conditions are met, as we discuss below.

\section{Discussion}

Even though the calculations presented in section \ref{sec:JWSTbroadbandfluxes} would suggest that population III stars might be detectable when viewed through a powerful gravitational lens, this only holds true if the most optimistic assumptions possible are adopted. Hence, no matter how one twists and turns, the prospects of detecting isolated population III stars in the foreseeable future appear bleak. However, contrary to previous claims, \cite{2006SSRv..123..485G} and \cite{2009MNRAS.399..639G}, the problem is not necessarily that population III stars are too faint for detection, since the magnification of a lens may actually lift them above the JWST detection threshold. The main obstacle is instead that the surface number densities for a sufficiently massive population III star is likely to be too low.

We have calculated the AB-magnitudes for population III stars using realistic models for the atmosphere and the HII region. However, our results are admittedly based on a number of simplifications, most of them made in the direction that would improve the prospects of detection. The spherically symmetric HII region that we have assumed is likely overestimating the flux, as there probably will be holes in the nebula. Recieving a 100h exposure time for a lensed field is probably overly optimistic so in reality the detection limit would be lower. The model selected is the model with highest SFR among the Trenti and Stiavelli models, thereby giving an overly generous estimate of the number of stars in a given area of the sky. The magnification is assumed to be exactly the magnification needed for observation. Finally our strategy of assuming that all population III stars attains masses of 300 M$_{\odot}$ probably overestimates the probability of detection. The only possibly pessimistic assumption is that of a negligible Ly$\alpha$ emission line.

\bibliographystyle{References}
{\small\bibliography{References}}

\providecommand{\href}[2]{#2}\begingroup\raggedright\begin{thebibliography}{10}

\bibitem{2004ARA&A..42...79B}
V.~{Bromm} and R.~B. {Larson}, {\it {The First Stars}},  {\em ARA\&A} {\bf 42}
  (Sept., 2004) 79--118,
  [\href{http://xxx.lanl.gov/abs/arXiv:astro-ph/0311019}{{\tt
  arXiv:astro-ph/0311019}}].

\bibitem{2009Natur.459...49B}
V.~{Bromm}, N.~{Yoshida}, L.~{Hernquist}, and C.~F. {McKee}, {\it {The
  formation of the first stars and galaxies}},  {\em Nature} {\bf 459} (May,
  2009) 49--54, [\href{http://xxx.lanl.gov/abs/0905.0929}{{\tt 0905.0929}}].

\bibitem{2010arXiv1006.1508C}
P.~C. {Clark}, S.~C.~O. {Glover}, R.~S. {Klessen}, and V.~{Bromm}, {\it
  {Gravitational fragmentation in turbulent primordial gas and the initial mass
  function of Population III stars}},  {\em ArXiv e-prints} (June, 2010)
  [\href{http://xxx.lanl.gov/abs/1006.1508}{{\tt 1006.1508}}].

\bibitem{2008ASPC..393..275S}
A.~{Stacy} and V.~{Bromm}, {\it {The Impact of Cosmic Rays on Population III
  Star Formation}},  in {\em New Horizons in Astronomy} ({A.~Frebel,
  J.~R.~Maund, J.~Shen, \& M.~H.~Siegel}, ed.), vol.~393 of {\em Astronomical
  Society of the Pacific Conference Series}, pp.~275--+, Aug., 2008.

\bibitem{2009ApJ...694..879T}
M.~{Trenti} and M.~{Stiavelli}, {\it {Formation Rates of Population III Stars
  and Chemical Enrichment of Halos during the Reionization Era}},  {\em ApJ}
  {\bf 694} (Apr., 2009) 879--892,
  [\href{http://xxx.lanl.gov/abs/0901.0711}{{\tt 0901.0711}}].

\bibitem{2001ApJ...552..464B}
V.~{Bromm}, R.~P. {Kudritzki}, and A.~{Loeb}, {\it {Generic Spectrum and
  Ionization Efficiency of a Heavy Initial Mass Function for the First Stars}},
   {\em ApJ} {\bf 552} (May, 2001) 464--472,
  [\href{http://xxx.lanl.gov/abs/arXiv:astro-ph/0007248}{{\tt
  arXiv:astro-ph/0007248}}].

\bibitem{2006SSRv..123..485G}
J.~P. {Gardner}, J.~C. {Mather}, M.~{Clampin}, R.~{Doyon}, M.~A. {Greenhouse},
  H.~B. {Hammel}, J.~B. {Hutchings}, P.~{Jakobsen}, S.~J. {Lilly}, K.~S.
  {Long}, J.~I. {Lunine}, M.~J. {McCaughrean}, M.~{Mountain}, J.~{Nella}, G.~H.
  {Rieke}, M.~J. {Rieke}, H.~{Rix}, E.~P. {Smith}, G.~{Sonneborn},
  M.~{Stiavelli}, H.~S. {Stockman}, R.~A. {Windhorst}, and G.~S. {Wright}, {\it
  {The James Webb Space Telescope}},  {\em SSR} {\bf 123} (Apr., 2006)
  485--606, [\href{http://xxx.lanl.gov/abs/arXiv:astro-ph/0606175}{{\tt
  arXiv:astro-ph/0606175}}].

\bibitem{2009MNRAS.399..639G}
T.~H. {Greif}, J.~L. {Johnson}, R.~S. {Klessen}, and V.~{Bromm}, {\it {The
  observational signature of the first HII regions}},  {\em MNRAS} {\bf 399}
  (Oct., 2009) 639--649, [\href{http://xxx.lanl.gov/abs/0905.1717}{{\tt
  0905.1717}}].

\bibitem{2002luml.conf..369H}
A.~{Heger}, S.~{Woosley}, I.~{Baraffe}, and T.~{Abel}, {\it {Evolution and
  Explosion of Very Massive Primordial Stars}},  in {\em Lighthouses of the
  Universe: The Most Luminous Celestial Objects and Their Use for Cosmology}
  ({M.~Gilfanov, R.~Sunyeav, \& E.~Churazov}, ed.), pp.~369--+, 2002.
\newblock \href{http://xxx.lanl.gov/abs/arXiv:astro-ph/0112059}{{\tt
  arXiv:astro-ph/0112059}}.

\bibitem{2002A&A...382...28S}
D.~{Schaerer}, {\it {On the properties of massive Population III stars and
  metal-free stellar populations}},  {\em A\&A} {\bf 382} (Jan., 2002) 28--42,
  [\href{http://xxx.lanl.gov/abs/arXiv:astro-ph/0110697}{{\tt
  arXiv:astro-ph/0110697}}].

\bibitem{1995ApJ...439..875H}
I.~{Hubeny} and T.~{Lanz}, {\it {Non-LTE line-blanketed model atmospheres of
  hot stars. 1: Hybrid complete linearization/accelerated lambda iteration
  method}},  {\em ApJ} {\bf 439} (Feb., 1995) 875--904.

\bibitem{1998PASP..110..761F}
G.~J. {Ferland}, K.~T. {Korista}, D.~A. {Verner}, J.~W. {Ferguson}, J.~B.
  {Kingdon}, and E.~M. {Verner}, {\it {CLOUDY 90: Numerical Simulation of
  Plasmas and Their Spectra}},  {\em PASP} {\bf 110} (July, 1998) 761--778.

\bibitem{2009ASPC..419..335J}
J.~L. {Johnson}, T.~H. {Greif}, V.~{Bromm}, R.~S. {Klessen}, and J.~{Ippolito},
  {\it {The First Galaxies: Signatures of the Initial Starburst}},  in {\em
  Astronomical Society of the Pacific Conference Series} ({S.~Jogee,
  I.~Marinova, L.~Hao, \& G.~A.~Blanc}, ed.), vol.~419 of {\em Astronomical
  Society of the Pacific Conference Series}, pp.~335--+, Dec., 2009.

\bibitem{2009ApJ...707L.102Z}
A.~{Zitrin}, T.~{Broadhurst}, Y.~{Rephaeli}, and S.~{Sadeh}, {\it {The Largest
  Gravitational Lens: MACS J0717.5+3745 (z = 0.546)}},  {\em ApJL} {\bf 707}
  (Dec., 2009) L102--L106, [\href{http://xxx.lanl.gov/abs/0907.4232}{{\tt
  0907.4232}}].

\bibitem{2010ApJ...717..257Z}
E.~{Zackrisson}, P.~{Scott}, C.~{Rydberg}, F.~{Iocco}, B.~{Edvardsson},
  G.~{{\"O}stlin}, S.~{Sivertsson}, A.~{Zitrin}, T.~{Broadhurst}, and
  P.~{Gondolo}, {\it {Finding High-redshift Dark Stars with the James Webb
  Space Telescope}},  {\em ApJ} {\bf 717} (July, 2010) 257--267,
  [\href{http://xxx.lanl.gov/abs/1002.3368}{{\tt 1002.3368}}].

\end{thebibliography}\endgroup

\end{document}